\documentclass[a4paper, oneside, twocolumn, notitlepage, 10pt]{extarticle_ecoc}
\usepackage{ecoc}

\usepackage{textcomp}
\usepackage{subcaption}
\usepackage{pgfplots}
\usepackage[dvipsnames]{xcolor} 
\usepgfplotslibrary{colorbrewer}
\pgfplotsset{width=10cm,compat=1.9}
\begin{document}
\raggedbottom
\selectlanguage{english}    


\title{Dual Line Coherent Detection}%


\author{
    Nelson Castro\textsuperscript{(1,$\ast$)}, Yiming Li\textsuperscript{(1,2)}, Mohammed Patel\textsuperscript{(1)}, Frank Smyth\textsuperscript{(3)}, Sergei Turitsyn\textsuperscript{(1)}, Andrew Ellis\textsuperscript{(1)}
}

\maketitle                  


\begin{strip}
    \begin{author_descr}

        \textsuperscript{(1)} AiPT, Aston University, Birmingham, UK,
        \textsuperscript{($\ast$)}\textcolor{blue}{\uline{n.castrosalgado@aston.ac.uk}}

        \textsuperscript{(2)} Shanghai Institute of Optics and Fine Mechanics, Chinese Academy of Sciences, Shanghai, China,

        \textsuperscript{(3)} Pilot Photonics, DCU Alpha Innovation Campus, Glasnevin, Dublin, Ireland

    \end{author_descr}
\end{strip}

\renewcommand\footnotemark{}
\renewcommand\footnoterule{}


\begin{strip}
    \begin{ecoc_abstract}
         We experimentally demonstrate dual-line coherent detection using an optical frequency comb local oscillator, enabling large 
         frequency offset tolerance 
         with minimal additional signal processing.
         The proposed method achieves
         200 GHz offset tolerance for 400 Gbit/s signals with low penalty, supporting uncooled, low-cost coherent transceivers. ©2026 The Author(s) 
    \end{ecoc_abstract}
\end{strip}


\section{Introduction}
It is widely acknowledged that coherent detection offers optimal receiver performance, enabling lower receiver sensitivities, 
lower launch powers
and access to the full optical field. It provides substantial advantages for data 
centre 
interconnects~\cite{zhou_beyond_2020}, satellite communications~\cite{walsh_demonstration_2022, Ren_weijie_intersatellite_2025} and access networks~\cite{Shahpari_coherent_access_2017}. However, the costs associated with digital signal processing and matching the signal and local oscillator (LO) wavelengths 
induce a reticence to deploy such solutions.
Innovations such as Coherent-Lite~\cite{Ye_coherent_lite_2025} address the former, whilst optical frequency comb (OFC)-based receivers~\cite{Zeng_frequency_offset_21,adib_tdm_pon_2022, Che_colorless_detection_2025}, the latter. 
Laser frequency drifts due to temperature and ageing exceed the highest reported receiver bandwidth \cite{Deakin_comb_receiver_2024}, making 
wavelength control  
necessary for conventional 
detection.
Even with sufficient receiver 
bandwidths, 
uncooled detection would demand a high excess sampling rate and enhanced frequency offset (FO) estimation, currently only supporting a range 
up to the symbol rate \cite{Lu_joint_2017}.
In contrast, an OFC-based LO improves FO range, allowing 
signal detection
with a receiver bandwidth 
close to
half the symbol rate, provided at least one comb line overlaps with the signal.
A
recently reported experiment
detected a 36~Gbaud signal with up to 36~GHz FO using a three-line OFC \cite{Zeng_frequency_offset_21}.

In this paper, we propose 
simplified 
signal processing, improving the FO range. 
We use pilot-based coarse FO estimation (FOE) 
to locate
two adjacent copies within the receiver bandwidth,
and stitch them with minimal digital signal processing (DSP).
Rather than
discarding non-complementary spectral portions
or adding adaptive equalisation, the available 
content from two 
signal copies is combined using quasi-static operations addressing comb-induced 
effects.
We show low-penalty recovery of a 64 Gbaud, 400 Gbit/s signal using two- and four-line LOs supporting FOs  of $\pm 68$~GHz and $\pm 100$~GHz, 
respectively.
\begin{figure*}[t]
    \centering
    \includegraphics[width=0.9\textwidth]{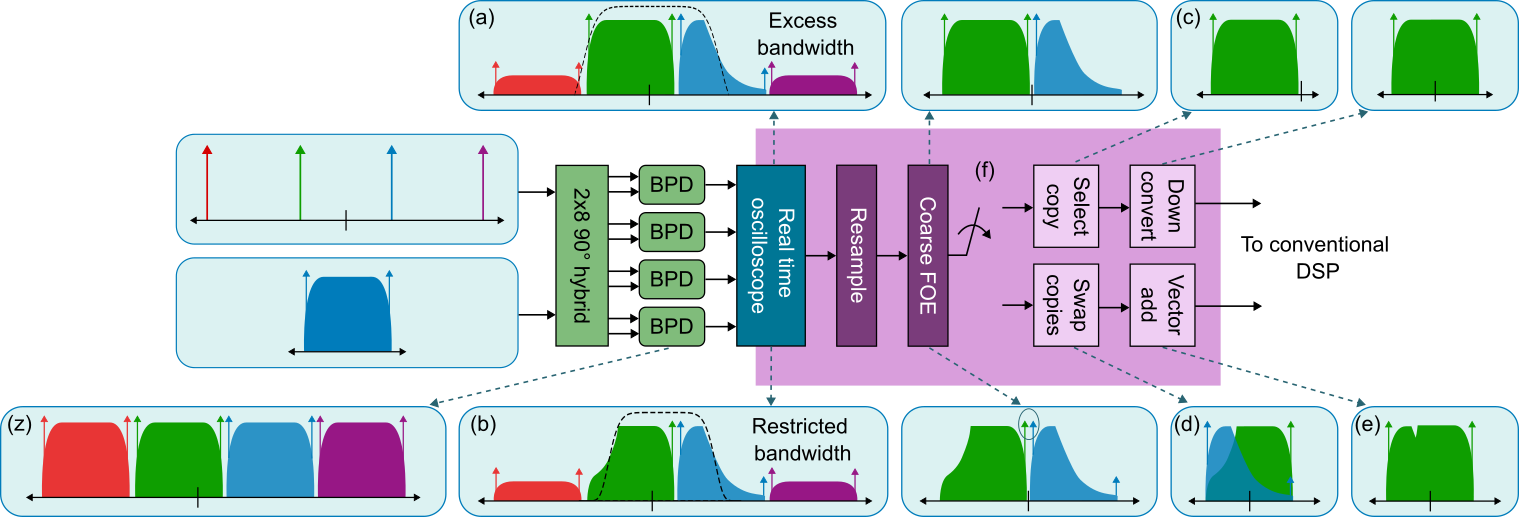}
    \caption{ 
    Operating principle of the proposed comb-based
    coherent detection. The received signal is coherently mixed with an OFC-based local oscillator, giving (z) one copy per comb line. 
    After (a) excess bandwidth and (b) limited bandwidth front-ends, some spectral copies are digitally available.
    In (a), an entire copy may be available, (c). In (b), detected fragments may be used to reconstruct the signal with appropriate frequency shifts (d) and weighted addition (e). Required DSP is shown in the purple box, (f). 
    }
    \label{fig:figure1}
\end{figure*}
\section{Operation Principle}
Combining the received field with an OFC-based LO
 generates multiple spectral copies within the receiver electrical bandwidth. With a suitable OFC spacing, such partial copies may be translated in frequency to reconstruct a complete image of the received signal, 
with FO tolerance scaling linearly with the number of OFC lines \cite{Zeng_frequency_offset_21}.
The proposed approach is shown in Fig. \ref{fig:figure1}, assuming a conventional dual-polarization coherent receiver.
A received signal is mixed with an OFC-based LO 
using a $90^\circ$ $2\times8$ optical hybrid and balanced photodetectors (BPDs).
The OFC spacing should be 
greater than or equal to the signal symbol rate,
and less than or equal to 
the receiver bandwidth.  
The result of the mixing process 
is shown in 
(a) and (b), 
illustrating one detected polarization
for excess and restricted receiver bandwidth, respectively.
In both cases,
mixing products resulting from the signal beating with the two nearest OFC lines will be available for signal processing, with the baseband electrical spectrum consisting of 
spectral copies of the signal with frequency shifts determined by the FO. 
With excess bandwidth, 
as in (a), there is potential for one copy to be complete. With restricted bandwidth, 
as in (b), a complete copy is only available  
when the received signal is closely aligned with one comb line, the general case being that two different fragments are detected, from which the received signal may be reconstructed.

In principle, spectral diversity allows straightforward detection of the 
copy positions
in the frequency domain, which can be used for coarse FOE. With the FO estimate, the signal may be down-converted and  
digital brick-wall filters used to discard parts of the spectral content, retaining only complementary portions to reconstruct the spectrum \cite{Zeng_frequency_offset_21}. Alternatively, shifted copies of the spectra could be processed in parallel, using full blind MIMO equalisation to reconstruct the spectrum \cite{adib_tdm_pon_2022,Othman_sliced_23}. 
To improve system performance in both bandwidth scenarios, we apply dual-copy processing (DC), where
we combine all the available spectral content of two copies \cite{adib_tdm_pon_2022}, performing simple frequency and phase shifting algorithms to align the copies, compensating for comb-induced frequency and phase offsets. 
This is followed by their weighted addition, 
which accounts for the static receiver frequency response
and allows 
spectral overlap. If  FOE suggests that a complete copy exists within the detected spectrum, selected-copy processing (SC) retains only the complete copy, applying brick-wall 
filtering and omitting vector addition.
This approach represents the simplest DSP requirement
to ensure negligible implementation penalty whilst enabling the maximum possible FO tolerance. 
The proposed dual-line DSP does not require any changes to the subsequent conventional stages, and avoids offloading coherent reconstruction operations to a power-hungry adaptive equalisation stage.

\section{Experimental Setup, Results and Discussion}

\begin{figure}[!b]
    \centering
    \includegraphics[width=1\columnwidth]{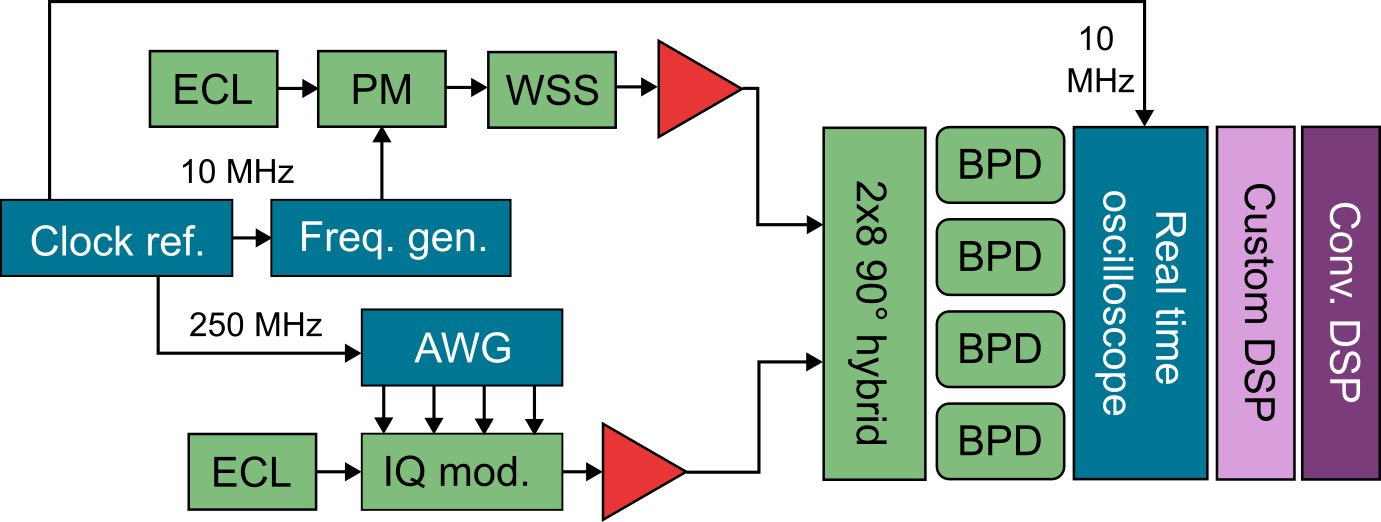}
    \caption{Schematic diagram of the experimental setup.}
    \label{fig:figure2}
\end{figure}

The experimental setup is depicted in Fig. \ref{fig:figure2}. At the transmitter, two 35280-bit pseudo-random sequences 
were generated and mapped onto 16-ary quadrature amplitude modulation (16QAM) symbols, with 1000 
4QAM
pilot symbols for frequency and phase synchronisation \textcolor{black}{inserted periodically at a rate of one pilot per nine symbols, amounting to a $10\%$ overhead} \cite{faruk_dsp_2017}. 
The in-phase and quadrature components of each polarisation were loaded onto a 
4-channel
arbitrary waveform generator with a sampling rate of 120 GSa/s, to give a 64 Gbaud signal.
Its 250 mV\textsubscript{pp} signals were amplified and drove the in-phase and quadrature (IQ) modulators of a \textcolor{black}{45}-GHz commercial 
transmitter.
\textcolor{black}{A $\sim$100-kHz linewidth
external cavity laser (ECL)} 
generated the optical carrier, 
and an erbium-doped fibre amplifier (EDFA) controlled the launch power of the optical signal.
At the receiver, 
a dual-line LO was generated 
from a second ECL, tuned to 193.1 THz, using a sine wave driven phase modulator (PM).
The 40-GHz bandwidth PM driven at 3-$V_{\pi}$ at 33 GHz produced a 7-line OFC (6 dB flatness).  A 
wavelength-selective switch (WSS)
extracted the comb lines for the LO.
For the dual-line LO, it retained only the 1st-order comb sidebands, located at \textcolor{black}{193.067 and 193.133 THz}. For the four-line LO, the 3rd-order sidebands at 193.001 and 193.199 THz were also retained, using the WSS to equalize the line powers.
For conventional single-line (SL) detection, only the seed laser wavelength was selected, and RF drive amplitude was reduced by 15 dB.
An EDFA amplified the resultant LO, ensuring  the three LO configurations had the same 13.2 dBm total optical power.
The carrier wavelength was detuned from that of the LO seed laser to induce FOs.
The incoming optical signal, with a typical OSNR of 30 dB, and the LO were mixed in a $90^\circ$ $2\times8$ optical hybrid with the outputs detected using 70 GHz BPDs.
The resulting electrical signals were captured with a 70-GHz 
real-time 
oscilloscope 
at a sampling rate of 200 GSa/s.
Synchronised clock signals for the RF and arbitrary waveform generators and the oscilloscope were derived from a single 10-MHz clock reference using a clock synthesizer board.
Ten traces of \textcolor{black}{$1\times10^{6}$} samples were captured for each wavelength detuning, which were processed offline in MATLAB. 
A diagram of the additions to the conventional DSP chain is shown in Fig. \ref{fig:figure1}(f).
The received signal was
\textcolor{black}{first resampled}.
Coarse FOE based on pilot tones inserted at the field edges at the transmitter was then employed \cite{Zeng_frequency_offset_21}, with tones placed 
at 
\textcolor{black}{$\pm 32.5 \, \mathrm{GHz}$}, just outside the signal spectrum.
The two tones located in between spectral copies of the signal, with a spacing of 1 GHz between them, were used for the coarse FOE.
The available copies were first frequency-shifted jointly by the FO estimate.
For SC processing,  
with a complete copy 
predicted from the FO estimate and the system parameters, the FO estimate was used to identify the copy closest to baseband and down-convert it.
For DC processing, both copies were first 
frequency-shifted in opposite directions by the OFC spacing, essentially swapping them in frequency, as shown in Fig. \ref{fig:figure1}(d). 
The phase of the copies was then aligned in the time domain, estimating the relative phase $\phi$ from the dot product of one copy with the complex conjugate of the other ($\mathrm{tan}(\phi) = \mathrm{Im}(\widetilde{E}_{1}^{*} \cdot \widetilde{E}_{2})/\mathrm{Re}(\widetilde{E}_{1}^{*} \cdot \widetilde{E}_{2})$).
Then, the 
copies were vectorially combined in the frequency domain, applying a raised-cosine crossfade profile across the overlap region, 
defined from the FO estimate and the available bandwidth. 
   \begin{figure*}
\centering

\newlength{\panelheight}
\setlength{\panelheight}{3.5cm}

\newlength{\rightgap}
\setlength{\rightgap}{0.15cm}

\newlength{\halfpanelheight}
\setlength{\halfpanelheight}{\dimexpr\panelheight/2-\rightgap/2\relax}

\newlength{\topaxisoffset}
\setlength{\topaxisoffset}{\dimexpr\halfpanelheight+\rightgap\relax}

\pgfplotsset{
    myaxis/.style={
        scale only axis,
        axis line style={black},
        tick align=outside,
        xtick pos=bottom,
        ytick pos=left,
        xlabel style={yshift=3pt},
        tick style={black},
        ymajorgrids=true,
        xmajorgrids=true,
        grid style=dashed,
        line width=0.9pt,
        clip marker paths=true,
        font=\small,
        legend style={
            draw=black,
            fill=white,
            font=\small,
            cells={anchor=west}
        }
    },
    sl/.style={
        Plum,
        very thick,
        mark=triangle*, mark options={
            draw=Plum,
            fill=Plum
        },
        mark size=2.4pt
    },
    dlsc/.style={
        OrangeRed,
        very thick,
        mark=square*,
        mark options={
            draw=OrangeRed,
            fill=OrangeRed
        },
        mark size=2.4pt
    },
    dldc/.style={
        color = PineGreen,
        very thick,
        mark=*, mark options={
            draw=PineGreen,
            fill=PineGreen},
        mark size=2.6pt
    },                          
    dlsc40/.style={
        WildStrawberry, mark=square*, mark options={
            draw=WildStrawberry,
            fill=WildStrawberry}, mark size=1.25pt
    },
    dldc40/.style={
        PineGreen,
        mark=*, mark options={
            draw=PineGreen,
            fill=PineGreen}, mark size=1.25pt
    },                          
    dlsc32/.style={
        WildStrawberry, mark=square*, mark options={
            draw=WildStrawberry,
            fill=WildStrawberry}, mark size=1.25pt
    },
    dldc32/.style={
        PineGreen,
        mark=*, mark options={
            draw=PineGreen,
            fill=PineGreen}, mark size=1.25pt
    }
}

\begin{subfigure}{0.4\textwidth}
\centering
\begin{tikzpicture}
\begin{semilogyaxis}[
    myaxis,
    width=\textwidth,
    height=\panelheight,
    y dir = reverse,
    ymin=1, ymax=6,
    ytick={1, 2,3,4,5,6},
    yticklabels={$-1$,$-2$,$-3$,$-4$,$-5$, $-6$},
    ylabel = log\textsubscript{10}(BER),
    xlabel={Frequency Offset (GHz)},
    xmin=-72, xmax=72,
    legend style={at={(0.9,0.95)}, 
    }
]

\addplot+[sl]
table[col sep=comma, x index=0, y index=1]
{Plots_data/log_ber_vs_offset_BW_Inf_GHz_single_line.csv};
\addlegendentry{SL}

\addplot+[dlsc]
table[col sep=comma, x index=0, y index=1]
{Plots_data/log_ber_vs_offset_BW_Inf_GHz_single_copy_equal_intervals.csv};
\addlegendentry{SC}

\addplot+[dldc]
table[col sep=comma, x index=0, y index=1]
{Plots_data/log_ber_vs_offset_BW_Inf_GHz_dual_copy_equal_intervals.csv};
\addlegendentry{DC}

\addplot[domain=-100:100, dashed, black, line width=0.5pt]{1.9031}node[pos=0.5, below,font=\scriptsize, xshift=-60pt] {$15\%$ OH FEC}; 

\addplot[
    only marks,
    mark=star,
    mark size=2.3pt,
    mark options={draw=white, line width=0.8pt},
] coordinates {(-4, 3.96)};

\node[anchor=south west, font=\small] at (rel axis cs:0.005,0.005) {(a)};

\node[
    anchor=north east,
    draw=black,
    line width=0.5pt,
    fill=white,
    inner sep=0.2pt
] at (rel axis cs:0.25,0.98)
{\includegraphics[width=1.1cm]{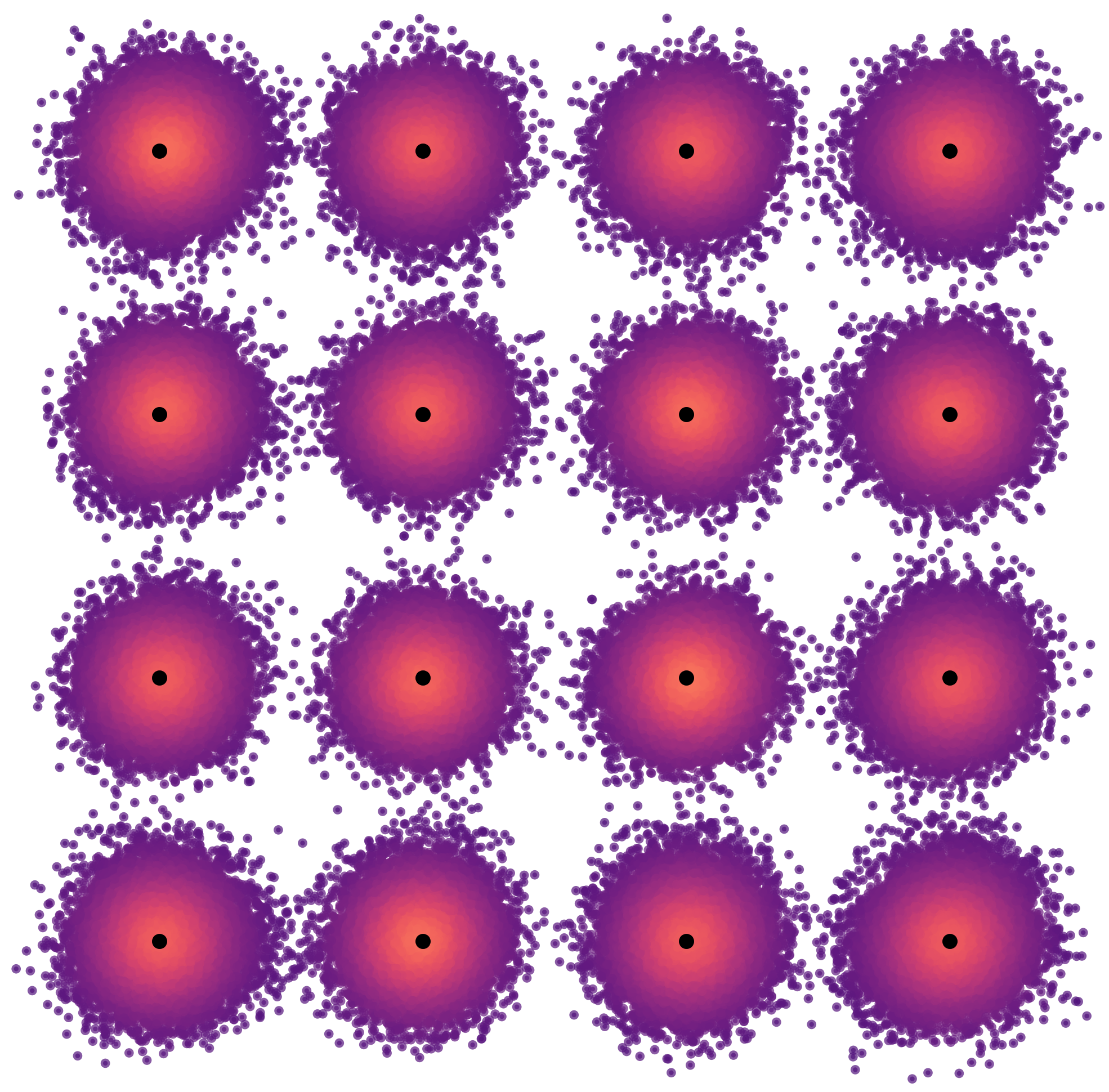}};

\end{semilogyaxis}
\end{tikzpicture}
\end{subfigure}
\hfill
\begin{subfigure}{0.51\textwidth}
\centering
\begin{tikzpicture}

\begin{axis}[
    myaxis,
    width=0.6\textwidth,
    height=0.85\halfpanelheight,
    at={(0,1.137\topaxisoffset)},
    anchor=south west,
    xmin=-62, xmax=62,
    y dir = reverse,
    ymin=0.3, ymax=5,
    ytick={1, 2,3,4,5, 6},
    yticklabels={$-1$,$-2$,$-3$,$-4$, $-5$, $-6$},
    xlabel={},
    ylabel={},
    legend style={at={(1.02,0.5)}, anchor=west}
]

\addplot+[dlsc40]
table[col sep=comma, x index=0, y index=1]
{Plots_data/log_ber_vs_offset_BW_40_GHz_single_copy.csv};
\addlegendentry{SC}

\addplot+[dldc40]
table[col sep=comma, x index=0, y index=1]
{Plots_data/log_ber_vs_offset_BW_40_GHz_dual_copy_equal_intervals.csv};
\addlegendentry{DC}

\addplot[domain=-100:100, dashed, black, line width=0.4pt]{1.9031}; 

\node[anchor=south west, font=\small] at (rel axis cs:0.005,0.005) {(b)};

\end{axis}

\begin{axis}[
    myaxis,
    width=0.6\textwidth,
    height=0.85\halfpanelheight,
    at={(0,0)},
    anchor=south west,
    xmin=-120, xmax=120,
    y dir = reverse,
    ymin=0.3, ymax=5,
    ytick={1, 2,3, 4, 5, 6},
    yticklabels={$-1$,$-2$,$-3$,$-4$,$-5$, $-6$},
    xlabel={Frequency Offset (GHz)},
    ylabel={},
    legend style={at={(1.01,0.5)}, anchor=west}
]
\node[anchor=south west, font=\small] at (rel axis cs:0.001,0.001) {(c)};

\addplot[domain=-120:120, dashed, black, line width=0.4pt]{1.9031}; 

\addplot+[dlsc32]
table[col sep=comma, x index=0, y index=1]
{Plots_data/log_ber_vs_offset_BW_32_GHz_single_copy.csv};

\addplot+[dldc32]
table[col sep=comma, x index=0, y index=1]
{Plots_data/log_ber_vs_offset_BW_32_GHz_dual_copy.csv};
\end{axis}

\end{tikzpicture}
\end{subfigure}

\caption{(a) Back-to-back performance against frequency offset for
single-line (SL) and the selected-copy (SC) and dual-copy (DC)
dual-line detection strategies with excess bandwidth. Inset shows X-pol. constellation at the $*$-marked point. 
Bandwidth-limited receiver performance for (b)  two comb lines and (c)  four comb lines with 40 GHz and 32 GHz bandwidth reception, respectively.
}
\label{fig:figure3}

\end{figure*}
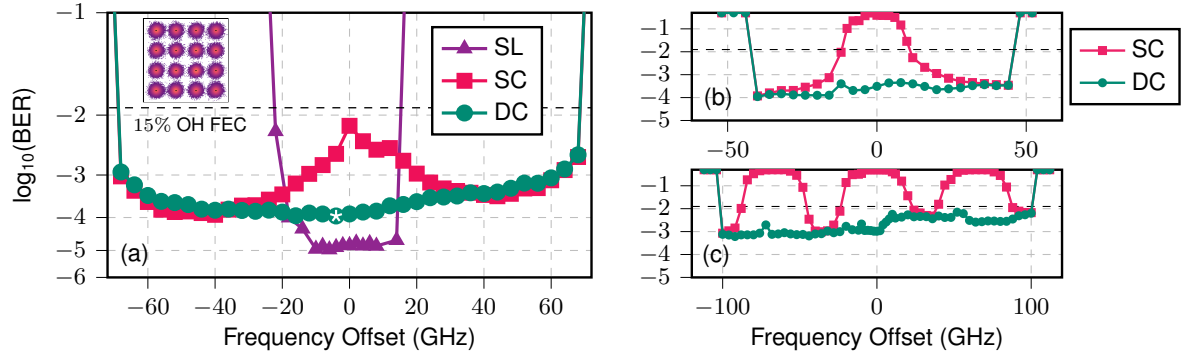
After the dual-line processing stages, a conventional coherent DSP chain was applied.
The in-phase and quadrature tributaries of each polarisation were first deskewed, followed by Gram-Schmidt orthonormalisation. 
The signals were then resampled to 2 samples per symbol and matched-filtered with a root-raised cosine (RRC) filter. 
The polarisations were demultiplexed, followed by timing recovery and frame synchronisation. Pilot-aided fine FO correction and carrier phase recovery were then applied. 
Lastly, pilot symbols were removed before bit-error counting.

Figure \ref{fig:figure3} shows the 
system bit-error rate (BER) performance against the FO between the transmitter laser and the comb seed laser. We consider a BER threshold of $1.25\times10^{-2}$ for $15\%$ overhead forward error correction (FEC) \cite{smith_leveraging_2017}. Results in Fig. \ref{fig:figure3}(a) consider an excess bandwidth system. The conventional SL 
system 
exhibits uniform performance with a 20-GHz FO tolerance ($30\%$ of the symbol rate). 
In contrast, the dual-line system has a much larger FO tolerance of 136 GHz ($212\%$ of the symbol rate).
The dual-line strategies SC and DC are shown in red and green, respectively.
DC offers enhanced performance over SC for small FO, because for such FOs both copies are degraded 
by the PD bandwidth and dual copy becomes advantageous. The depth of the SC performance dip, and hence the need for DC processing, is determined by the extent of the excess receiver bandwidth, which here is $110\%$. In contrast, such frequency response results in small penalties at the highest FO for DC processing, potentially due to the influence of noise surrounding the spectrum of the smaller copy, degrading the overall SNR.

In light of the importance of receiver bandwidth, we investigated the system performance for other PD bandwidths, emulated digitally by applying a low-pass Super Gaussian filter of order 4. The performance of 
DC and SC
is shown for 
a bandwidth of 40 GHz and two comb lines (Fig. \ref{fig:figure3}(b)) and 32 GHz and four comb lines (Fig. \ref{fig:figure3}(c)). 
In the SC case, the bandwidth restriction substantially increases the penalty for cases where no complete copy is available.  Conversely, DC enables the system to perform well across the expected FO range, successfully reconstructing the signal spectrum from the partial copies over FO ranges of 84 and 200 GHz for 2- and 4-line LOs, respectively. 

A
modest penalty
is observed for the OFC-based LO compared to the conventional system, even with the imposed bandwidth restrictions.
Since the system is operated 
in the high optical signal to noise ratio (OSNR) regime to fully illustrate frequency dependent effects, the dominant noise source is LO shot noise.
As the total LO power is split between comb lines, a degradation in SNR of 1.5 dB (2-line OFC) and 3 dB (4-line OFC) is expected from the increased shot noise of each line. This compares well to observed shot noise penalties of 1.7 dB and 3.4 dB
for the 2- and 4-line cases,
calculated by converting the lowest recorded BER to an effective SNR \cite{Ellis_performance_17}. Greatly reduced penalties may be expected in the OSNR-limited regime, \textcolor{black}{where} both signal and dominant noise powers scale equally with the number of comb lines.

\section{Conclusions}
We have presented an OFC-based coherent detection method that greatly relaxes the required FO tolerance.
The system achieves the 
expected FO tolerance, with 
SNR penalties close to LO shot noise predictions.
It supports a data rate of 400 Gbit/s, with FO tolerances of 200 GHz for a 4-line LO 32-GHz receiver and 136 GHz for a 2-line LO 70-GHz receiver.
\section{Acknowledgements}
This work was supported by the UK EPSRC grant Advanced Optical Frequency Comb Technologies and Applications (EP/W002868/1).

\defbibnote{myprenote}{}
\printbibliography[prenote=myprenote]

\vspace{-4mm}

\end{document}